\documentclass[%
 reprint,
 amsmath,amssymb,
 aps,
]{revtex4-2}

\usepackage{graphicx, verbatim, multirow, enumitem}
\usepackage[usestackEOL]{stackengine}
\usepackage{dcolumn}
\usepackage{bm}
\newcommand{\subfigimg}[3][,]{%
  \setbox1=\hbox{\includegraphics[#1]{#3}}
  \leavevmode\rlap{\usebox1}
  \rlap{\hspace*{10pt}\raisebox{\dimexpr\ht1-2\baselineskip}{#2}}
  \phantom{\usebox1}
}
\usepackage{dcolumn}
\usepackage{bm}

\begin{document}

\preprint{APS/123-QED}

\title{A Master Equation for Power Laws}

\author{Sabin Roman$^{1, 2}$ and Francesco Bertolotti$^{3}$}
 \affiliation{$^{1}$Centre for the Study of Existential Risk, University of Cambridge\\
 $^{2}$Odyssean Institute\\
 $^{3}$Carlo Cattaneo University LIUC}

\date{\today}

\begin{abstract}
We propose a new mechanism for generating power laws. Starting from a random walk, we first outline a simple derivation of the Fokker-Planck equation. By analogy, starting from a certain Markov chain, we derive a master equation for power laws that describes how the number of cascades changes over time (cascades are consecutive transitions that end when the initial state is reached). The partial differential equation has a closed form solution which gives an explicit dependence of the number of cascades on their size and on time. Furthermore, the power law solution has a natural cut-off, a feature often seen in empirical data. This is due to the finite size a cascade can have in a finite time horizon. The derivation of the equation provides a justification for an exponent equal to 2, which agrees well with several empirical distributions, including Richardson's law on the size and frequency of deadly conflicts. Nevertheless, the equation can be solved for any exponent value. In addition, we propose an urn model where the number of consecutive ball extractions follows a power law. In all cases, the power law is manifest over the entire range of cascade sizes, as shown through log-log plots in the frequency and rank distributions.
\end{abstract}

\maketitle


\section{Introduction}

A power law is a non-linear relationship between two quantities $x$ and $y$ that can be modelled generically by the following formula: $y = ax^k$, where $k$ and $a$ are constants, respectively the exponent of the power law,  and the width of the scaling relationship. The power law is scale-invariant relationship $f(xc) \propto f(x)$ that holds for any value of $c$. Graphically, it implies that the curve describing the relationship between $x$ and $y$ maintain its shape under any possible dilatation. Moreover, power laws can be represented as straight lines on a log-log plot, called the signature of the power law.

The signature can be employed to analyze empirical data by comparing the distribution of the data on a log-log plot with the best fitting power law. Likely, this representation was first introduced by J. C. Wills in 1922 to plot the distribution of the number of species in a genus \cite{Newman2005}. Many log-normal relationships appear as power laws when plotted for small ranges. As a consequence, to assert that a relationship between two variables is a power law, it should hold for at least two orders of magnitudes \cite{Stumpf2012}.

In the next section, we enumerate some representative examples of power law distributions from physics, biology, ecology and social sciences. Furthermore, we list the most well-known mechanisms for generating power laws. In the third section, we proceed to propose a Markov process model for generating power laws. First, we outline a derivation of the Fokker-Planck equation by starting from a random walk and taking a continuum limit from a discrete state space. This is a well-known approach for deriving the Fokker-Planck equation. We provide our own treatment of this procedure by changing the difference operator $\Delta$ acting on discrete functions to the differential operator $d$ that acts on real variables. We then propose a Markov chain that displays a power law in its long-term equilibrium distribution. Based on this and in analogy with the Fokker-Planck equation derivation, we obtain a master equation for power laws that we simplify to an analytically tractable form. Surprisingly, the solution can be written in closed-form. Afterwards, in the fourth section, we propose an urn model where the number of cascades (consecutive balls drawn) scales according to a power law. This improves on an urn model by Brunk \cite{Brunk2000} that display a power law relationship only in a limited range. Finally, we summarize our results in the conclusion.

\section{Background}

The main challenge regarding power law distributions is understanding their origin. Broadly speaking, there are two ways power laws can emerge: probability transformations and generative processes. Probability transformations derive power law distributions from other distributions. Generative processes are algorithms that create new distributions. Here, we provide more details on generative processes, because our own work is of this nature. To the best of our knowledge, there does not exist a comprehensive review of all the power law distribution generative models (even if \cite{Newman2005} finds a large set), and it is not in the scope of this paper to provide one. 

There are three main strategies to obtain a power law distribution by transforming an existing probability distribution: combining exponential distributions, inverting quantities or looking at the extreme values of distributions. 

In the first case, the transformation consists of substituting the variable of an exponential distribution with another exponential distribution. The resulting distribution follows a power law, in which parameters depend on the original exponential distribution. The distribution of sizes of populations that grow exponentially (i.e., with infinite carrying capacity) and can suddenly become extinct at any time step with the same probability follows a power law that can be derived from this model \cite{Reed2001}.

The second case requires the inversion of the quantities of the probability distribution. If these quantities pass through zero, the transformation results in a power law distribution with exponent equal to $2$, resulting from the derivation of an inverted variable. The most notable example of a power law distribution obtained in this way occurs in the paramagnetic phase of the Ising model \cite{Jan1999, Krasnytska2021}.

The third way originates from the extreme value theorems which provide results on the asymptotic behavior of the extreme realizations. The Pickands–Balkema–de Haan theorem \cite{Balkema2007, Pickands1975} states that the conditional distribution of a random variable above a certain threshold tends to a Pareto distribution when the threshold tends to infinity. Therefore, it provides a rationale for the widespread observation of Pareto or power law behavior, since it is the limiting behavior of large events for a whole class of probability distributions \cite{Alfarano2010}. Analogously, the Fisher–Tippett–Gnedenko theorem \cite{Fisher1928, Gnedenko1943} describes the possible distributions that the maximum value can have; this can shed light in cases where a simple power law does not appear from the application of the Pickands–Balkema–de Haan theorem \cite{Sornette2006}.

In addition, we identified in the literature at least four notable generative processes for power law distributions: phase transitions (along with self-organized criticality), random walks, sample-space reduction processes, and the Yule processes. 

The idea underlying phase transitions and critical phenomena is simple but powerful. In systems governed by a single length scale, the scale can diverge in some given conditions, giving birth to scale-free systems in which quantities are distributed as a power law. The precise point at which the length-scale diverges is called a critical continuous point, or phase transition. Nevertheless, it is unlikely that the parameters that regulate the phase-transition of a real-world system happen to fall on that specific value. So, the existence of critical phenomena is not enough to explain the presence of power law distributions in many natural and social systems. But some systems appear to self-organize to lay close to critical points, independently from the initial conditions. This phenomenon, called self-organized criticality, can be exemplified by the classic sandpile model \cite{Bak1987}, which generates avalanches whose sizes are distributed according to a power law. Other examples of how self-organized criticality generates power law distributions includes earthquakes \cite{Olami1992}, neuronal avalanches \cite{Beggs2003}, and forest fires \cite{Newman2005, McKenzie2012}. However, the emergence and applicability of self-organized criticality is debatable \cite{frigg2003self}.

Random walks are a succession of randomly-generated steps on a given space \cite{Hughes1995}. The statistical properties of random walks tend toward universal distributions when the steps are independent of each other, and their number grows unbounded \cite{Vignat2006}. As a consequence, they have the ability to generate power law distributions even when the underlying rules are very simple. One of the most famous examples (and likely the simplest) relates to the first-return time. It can be proved that, under specific conditions, the distributions of the first-return times follows a power law \cite{Vignat2006, Newman2005}.

The sample-space reduction process models history-dependent systems, such as the formation of sentences, whose number of possible different meanings is progressively reduced the more words are added to it. Corominas-Murtra et al. demonstrate that a power law distribution can be generated from such a process \cite{Corominas-Murtra2015}. More specifically, Zipf's law necessarily emerges from the reduction over time of the number of possible states in which a system can be, as a consequence of symmetry breaking in random sampling processes \cite{Corominas-Murtra2015} or from dependency structures in component systems \cite{Mazzolini2018}. The reduction space process explains some well-known domain-specific power law generation models, such as preferential attachment. 

The Yule process is a generative model developed by G. Udny Yule to explain the power law distribution of the number of species in taxonomic groups \cite{E.1925}. It derives from simple rules. Firstly, starting from a pool of groups, the probability for a group to increase by one is proportional to the number of its elements. Secondly, at any speciation event, there is a possibility of generating a new specie belonging to a brand-new taxonomic group. The model is a simplification of reality since it ignores extinction events. Nevertheless, it has been customized to explain the emergence of power laws in other systems such as city sizes \cite{Simon1955} or paper citations \cite{Krapivsky2000}.

Finally, other generative processes include highly optimized tolerance \cite{Carlson1999, Carlson2000}, the coherent noise model of biological extinction \cite{Sneppen1997}, the repeated fragmentation model of fixed length elements \cite{Newman2005}, the dynamic of times between records in a random process \cite{Sibani1993}, the Hawkes processes \cite{Kanazawa2020} and out-from-criticality feedback \cite{Schulman2021}.

A master equation for power laws has been previously obtained in the literature \cite{biro2017dynamical,biro2018unidirectional, biro2019entropic, gere2021wealth, biro2021transient} but only considers stationary distributions or transient dynamics, while our results give a non-stationary solution \cite{inacio2022comments}. Also, our choice of transition rates has not been considered before.

To the best of our knowledge, the above examples and mechanisms for generating power laws do not generally address the question: How do power laws change (or are preserved) over time? Answering this question requires a way to specify the time evolution of the power law distribution. Furthermore, in many cases, empirical power laws have a cut-off point above which the relationship no longer holds \cite{Newman2005} and this is has not been captured by generative models, which are realistic only in specific value ranges. The next section provides a mechanism by which an equation dictating the time-evolution of power laws can be obtained. In addition, the solution displays a natural cut-off beyond a certain time horizon.

\begin{figure*}[t]
  \centering
\includegraphics[width=\textwidth]{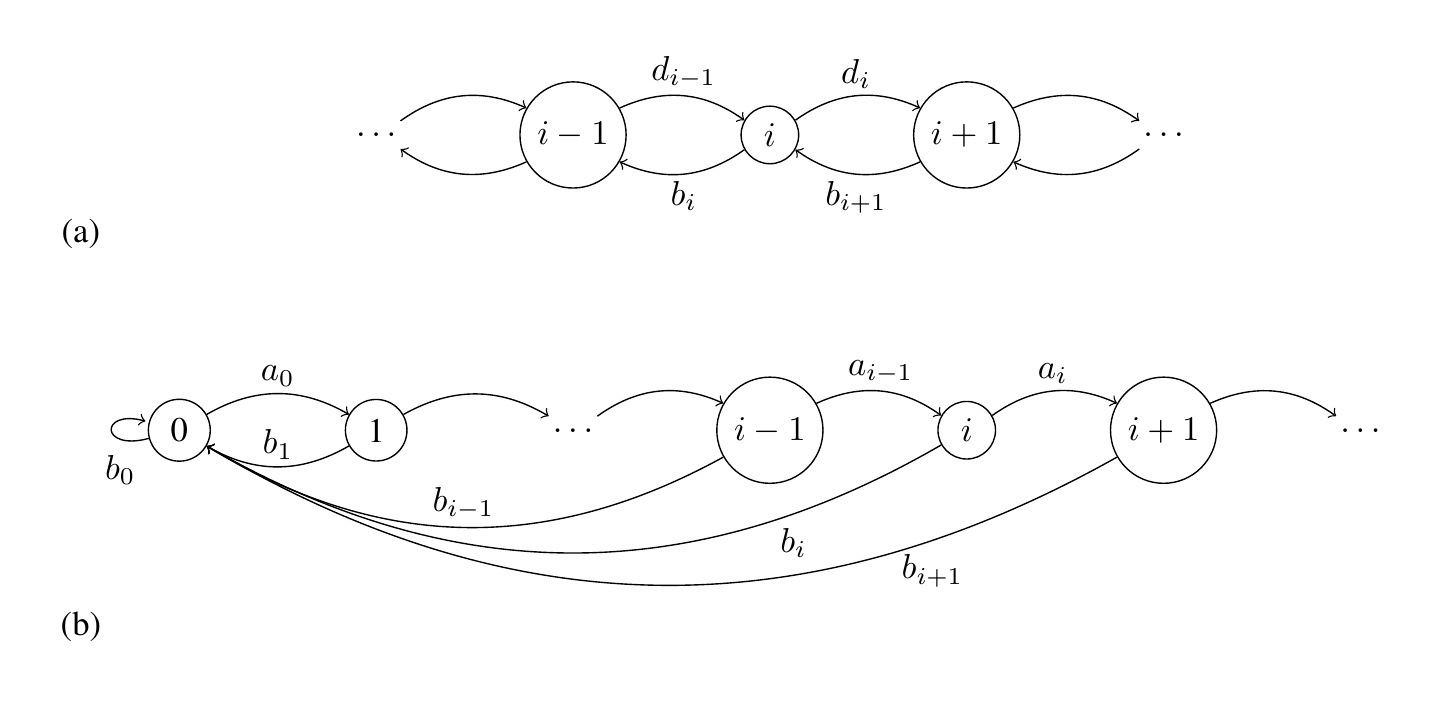}
\caption{Discrete state Markov chains for: (a) a continuous time random walk, with multi-step transition probabilities satisfying equation \eqref{eq:rweq}, (b) a continuous time process, with multi-step transition probabilities satisfying equation \eqref{eq:pldisc}. The stationary distribution of the embedded Markov chain is a power law, as shown in derivation \eqref{eq:plder}.}
\label{fig:0}
\end{figure*}

\section{Markov process models}

Below we present an outline for a well-known derivation of the Fokker-Planck equation starting from a discrete Markov chain and taking a continuum limit. By analogy, we introduce a discrete Markov chain that has a power law stationary distribution and then we derive a master equation for power laws.

\subsection{Fokker-Planck equation}

The transition probabilities for a 1D random walk are:
\begin{equation}
\begin{aligned}
    P_{i, i+1} &= p\\
    P_{i, i-1} &= 1-p = q
\end{aligned}
\end{equation}

The probability for $x$ jumps to the right and $n-x$ jumps to the left is:
\begin{equation}
\begin{aligned}
    P(x, n) &= \binom{n}{x} p^{x} q^{n-x}\\
            &\sim \cfrac{1}{\sqrt{2\pi npq}}\exp\left[{-\cfrac{(x-np)^{2}}{2npq}}\right]
\end{aligned}
\end{equation}

where in the limit of large enough $n$ we can can approximate the binomial distribution with a Gaussian. Let $i, s$ be arbitrary states. For a continuous time Markov chain we write $p_{i}(t) = P_{si}(0; t)$. For the Markov chain in Fig. \ref{fig:0}(a), we can write down the following equation for $p_{i}(t)$:
\begin{equation}
\begin{aligned}
    \frac{dp_{i}(t)}{dt} &= d_{i-1}p_{i-1} + b_{i+1}p_{i+1} - d_{i}p_{i} - b_{i}p_{i}\\
    &= -(a_{i+1}p_{i+1}-a_{i}p_{i})\\ &\;\;\;\;\;  + d_{i-1}p_{i-1}-2d_{i}p_{i} + d_{i+1}p_{i+1}
\end{aligned}
\label{eq:rweq}
\end{equation}

where $b_{i}, d_{i}$ are the rates at which the process leaves state $i$ to state $i-1$, respectively to state $i+1$. Furthermore, we define $a_{i} = d_{i}-b_{i}$. Using the forward and backward difference operators $\Delta x_{n} = x_{n+1}-x_{n}$ and $\nabla x_{n} = x_{n}-x_{n-1}$, the above equation can be written:
\begin{equation}
\begin{aligned}
    \frac{dp_{i}(t)}{dt} &= -\Delta (a_{i}p_{i}(t)) + \Delta\nabla(d_{i}p_{i}(t))\\
    &\downarrow\\
    \frac{\partial p(x, t)}{\partial t} &= -\frac{\partial}{\partial x}(a(x)p(x, t)) + \frac{\partial^{2}}{\partial x^{2}}(d(x)p(x, t))
\end{aligned}
\end{equation}

which in the continuous limit gives the Fokker-Planck equation. If $a(x) = 0$ and $d(x) = D$ where $D$ is constant, then the solution with initial condition $p(0, x) = \delta(x-x_{0})$ is given by:
\begin{equation}
    p(x, t) = \cfrac{1}{\sqrt{4\pi Dt}}\exp\left[-\cfrac{(x-x_{0})^{2}}{4Dt}\right]
\label{eq:fpsol}
\end{equation}

\begin{figure*}[t]
  \centering
  \begin{tabular}{@{}p{0.5\linewidth}@{\quad}p{0.5\linewidth}@{}}
    \subfigimg[width=\linewidth]{\hspace{4cm}(a)}{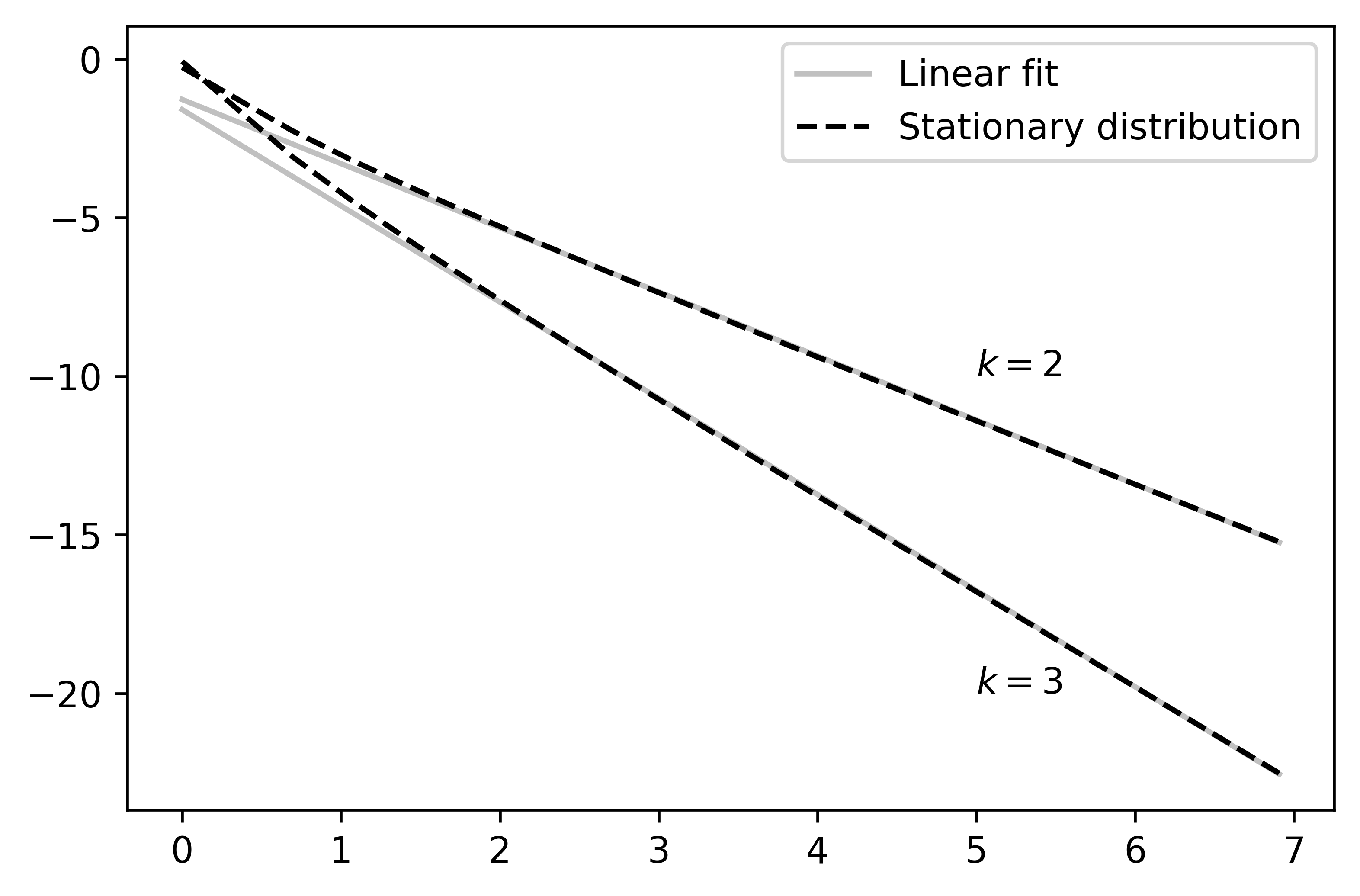} &
    \subfigimg[width=\linewidth]{\hspace{4cm}(b)}{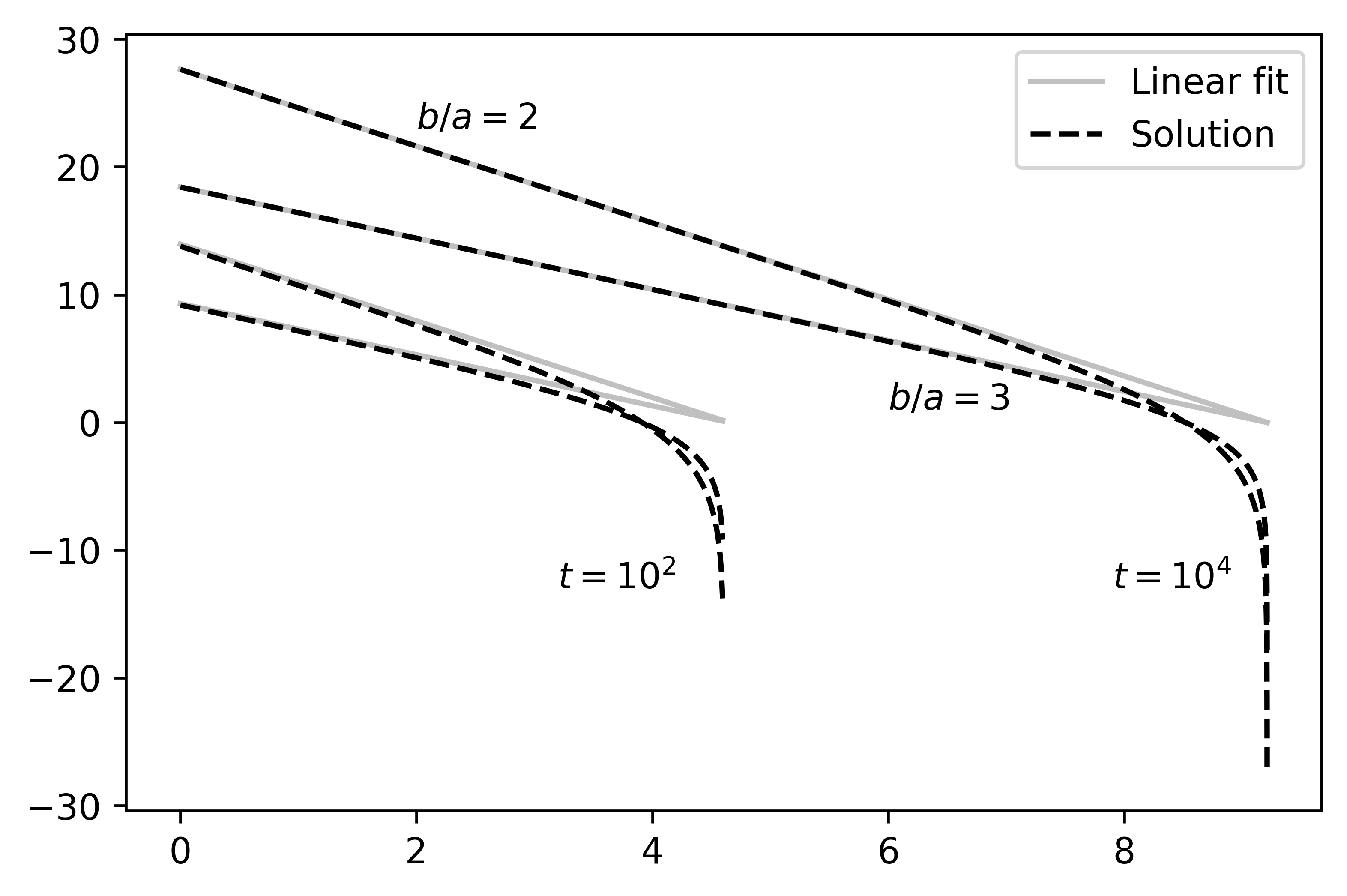}
  \end{tabular}
\caption{Log-log plots of: (a) The stationary distribution (dotted black line) of the Markov chain with transition probabilities \eqref{eq:pltp} for $k = 2, 3$, along with a linear fit (solid grey line). (b) Solution \eqref{eq:csol} (dotted black line) to equation \eqref{eq:pleq}  for $b/a = 2, 3$ up to times $t = 10^{2}, 10^{4}$, along with linear fits (solid grey line).}
\label{fig:1}
\end{figure*}

\subsection{Power law equation}
\label{subsec:pleq}

We consider the Markov chain with the following transition matrix:
\begin{equation}
\begin{aligned}
    P_{i, i+1} &= \exp\left(-\frac{k}{i+1}\right)\\
    P_{i, 0} &= 1-\exp\left(-\frac{k}{i+1}\right)
\end{aligned}
\label{eq:pltp}
\end{equation}
The dominant contribution to the long-term equilibrium distribution is given by:
\begin{equation}
\begin{aligned}
    \pi(x) &\simeq \Pi_{i \leq x} P_{i, i+1}\\
           &= \exp(-k \sum_{i} 1/i)\\
           &\sim \exp(-k \ln{x})\\
           &= x^{-k}
\end{aligned}
\label{eq:plder}
\end{equation}

In Fig. \ref{fig:1}(a) we compare numerical solutions for the equilibrium distribution with the approximation \eqref{eq:plder}, showing that the power law approximation is highly accurate. The Markov chain given by \eqref{eq:pltp} is the embedded chain of the continuous time Markov process in Fig. \ref{fig:0}(b). Using the same notation as in the previous section, the probability $p_{i}(t)$ satisfies the following differential equation:
\begin{equation}
    \frac{dp_{i}(t)}{dt} = a_{i-1}p_{i-1}-a_{i}p_{i} - b_{i}p_{i}
\label{eq:pldisc}
\end{equation}

We can write \eqref{eq:pldisc} using the backward difference operator. If we let the space index go from discrete to continuous, such as in the previous section where $i \rightarrow x$, then equation \eqref{eq:pldisc} becomes:
\begin{equation}
\begin{aligned}
    \frac{dp_{i}(t)}{dt} &= -\nabla (a_{i}p_{i}(t)) - b_{i}p_{i}(t)\\
    &\downarrow\\
    \frac{\partial p(x, t)}{\partial t} &= -\frac{\partial}{\partial x}(a(x)p(x, t)) - b(x)p(x)
\end{aligned}
\label{eq:pleq}
\end{equation}

which gives:
\begin{equation}
    \frac{\partial p(x, t)}{\partial t} = -a(x)\frac{\partial p(x, t)}{\partial x} - (a'(x) + b(x))p(x)
\label{eq:aux}
\end{equation}

where $a(x) =  a e^{-k/x}, b(x) = b (1-e^{-k/x})$ and $a, b$ are constants. By Taylor expanding up to $1/x$, we get $a(x) \simeq a - \frac{ak}{x}$ and  $a'(x) + b(x) \simeq \frac{bk}{x}$. Without loss of generality we can relabel $bk$ as $b$ and the equation reduces to:
\begin{equation}
\frac{\partial p(x, t)}{\partial t} = -a\frac{\partial p(x, t)}{\partial x} -  \frac{b}{x}p(x, t)
\label{eq:plsim}
\end{equation}

The term containing $a/x$ can be neglected because the probability $p(x, t)$ depends on a power of $x$ (see below), and so each term on the right-hand side is of \eqref{eq:plsim} is of the same order in $x$.

Let $c(x, t)$ be the unnormalized solution of equation \eqref{eq:plsim}. The solution  $c(x, t)$ has a closed form expression for $x > 0$ and $x < at$:
\begin{equation}
c(x, t) = c_{0}\left(\frac{at}{x} - 1\right)^{b/a}
\label{eq:csol}
\end{equation}

where $c(x, 0) = c_{0}$. In Fig.\ref{fig:1}(b) we plot the solution \eqref{eq:csol} for different values of $b/a$ and see a power law behaviour over a range of scales that increases over time. The tail of the solution falls off abruptly just as we see in many real-world examples of power law distributions \cite{Newman2005}.

\begin{figure*}[t]
  \centering
  \begin{tabular}{@{}p{0.5\linewidth}@{\quad}p{0.5\linewidth}@{}}
    \subfigimg[width=\linewidth]{\hspace{4cm}(a)}{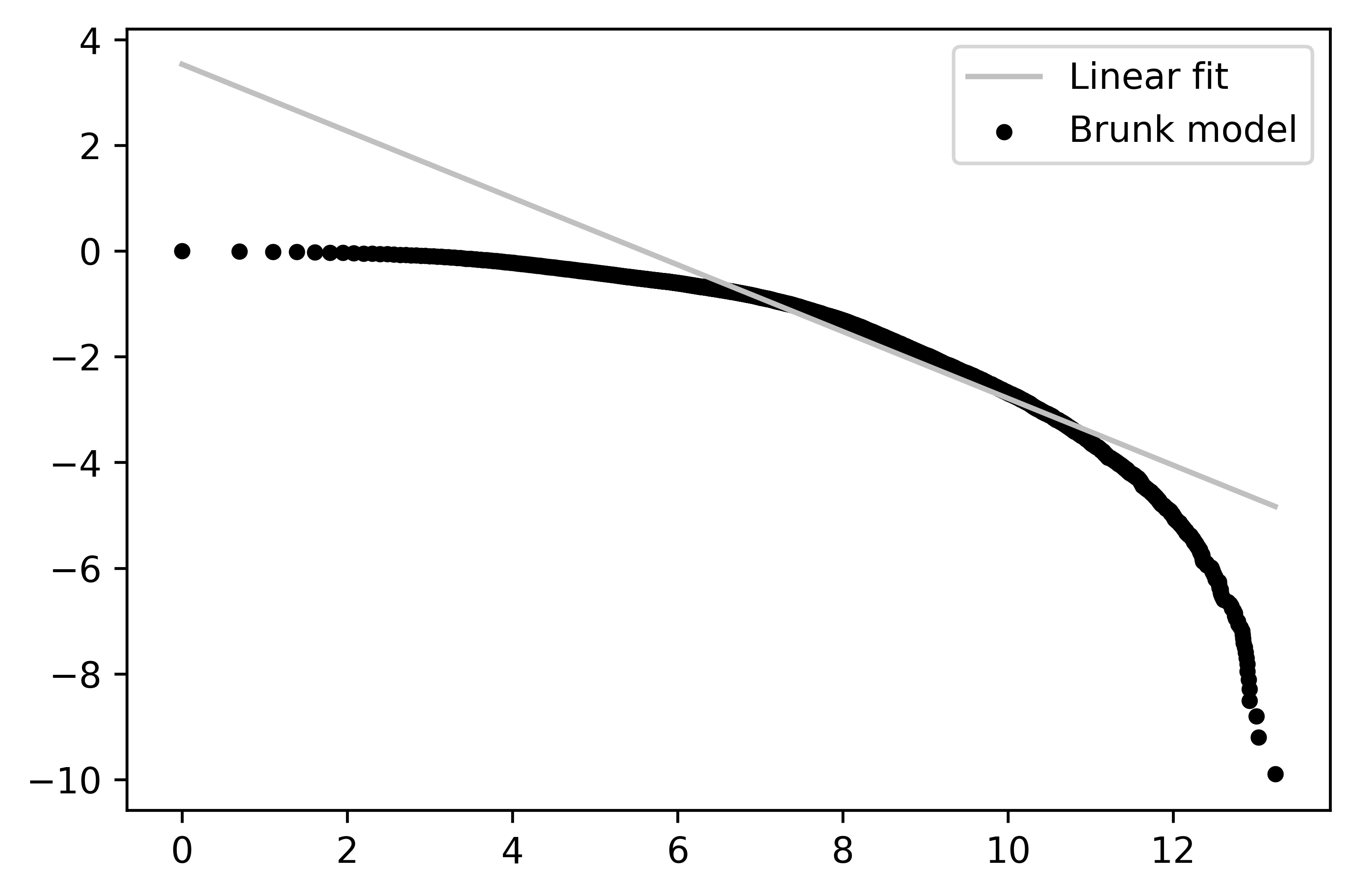} &
    \subfigimg[width=\linewidth]{\hspace{4cm}(b)}{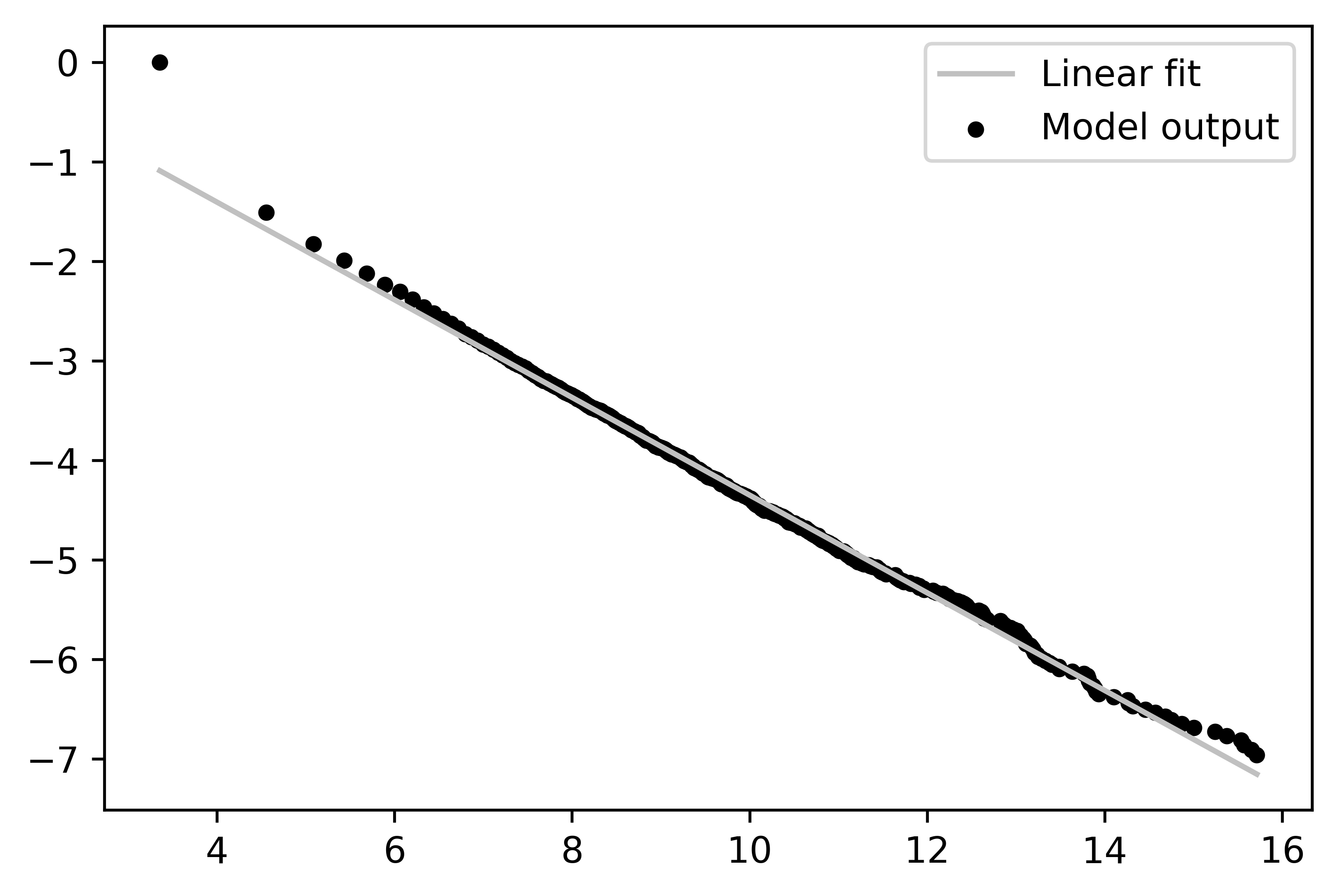}
  \end{tabular}
\caption{Log-log plots of the rank distribution of: (a) Cascade sizes (black dots) in Brunk's model, showing a narrow range where a power law is observed, along with a linear fit (solid grey line). (b) Cascade sizes (black dots) in the model we propose with exponent $k = 1.5$, along with a linear fit (solid grey line). The line of best fit has slope $0.49$, in good agreement with the theoretical value of $k-1 = 0.5$.}
\label{fig:2}
\end{figure*}

Equation \eqref{eq:plsim} posits that cascades (consecutive transitions) occur over time in inverse proportion to their size. Furthermore, there is a constant rate $a$ at which they can increase in size. For a finite time horizon $t$, there is a maximum size that a cascade can reach, namely $at$, as can be seen in the solution \eqref{eq:csol}.

We can get additional insights from equation \eqref{eq:aux} if we consider the second order Taylor expansion of  $a'(x) + b(x) \simeq \frac{bk}{x} + \frac{2ak-bk^{2}}{2 x^{2}}$. If $k = 2 a/b$ then the $1/x^{2}$ term cancels, and this implies an exponent of $2$ in the solution \eqref{eq:csol}. This is not necessary in deriving equation \eqref{eq:plsim} but can serve as a rationale for the experimental observation of power laws with exponent $2$.

\section{Urn model}

Building on the insights of the previous section, we propose an urn model that display power law behaviour. We can approximate the probability \eqref{eq:pldisc} to move from $i$ to $i+1$ state as:
\begin{equation}
    \exp\left(-\frac{k}{i+1}\right) \simeq 1-\frac{k}{i+1}
\end{equation}
for large enough $i$. Hence, if an adequate value of $k$ is chosen, then we can expect to see a power law emerging even in a simple model.

\cite{Brunk2000} proposes a model claiming to exhibit cascades that occur with frequencies inversely proportional to the cascade size, and provides graphs showing that a power law distribution emerges. We attempted to reproduce the results of the paper but we obtained an exponential distribution instead of a power law. A power law relationship is seen to occur in a narrow range of scales that the graphs in the article use, see Fig. \ref{fig:2}(a). Furthermore, we can analytically prove that the distribution that emerges is an exponential distribution and not a power law. The article \cite{Brunk2000} has been cited in recent years, and we consider it important to highlight the limitations of the model.

Assume the urn contains $N$ balls in total, with $K$ balls black and the rest white. The model has the following steps:
\begin{enumerate}
    \item A ball is drawn, if it is black then the next ball is drawn, and the process is repeated until a white ball is drawn. This constitutes a cascade within the model. Black balls are drawn without replacement.
    \item Each time a white ball is drawn the number of black balls is increased by a fraction $g$, while the number of white balls stay the same. If a white ball is drawn, then we repeat from step 1.
\end{enumerate}
The probability of that a cascade of size $n$ occurs, i.e., $n$ black balls are drawn followed by a white ball, is:
\begin{equation}
\begin{aligned}
P(S=n) &= \frac{K}{N}\cdot\frac{K-1}{N-1}\cdot\ldots\cdot\frac{K-n+1}{N-n+1}\cdot\frac{N-K}{N-n}\\
&=\frac{K!}{N!}\cdot\frac{(N-n)!}{(K-n)!}\cdot\frac{N-K}{N-n}\\
&\approx \frac{K^{K+1/2}}{N^{N+1/2}}\cdot\frac{(N-n)^{N-n+1/2}}{(K-n)^{K-n+1/2}}\cdot\frac{N-n-(K-n)}{N-n}\\
&=\left(1-\frac{n}{N}\right)^{N+1/2}\left(1-\frac{n}{K}\right)^{-(K+1/2)}\\ &\;\;\;\;\left(\frac{K-n}{N-n}\right)^{n}\left(1-\frac{K-n}{N-n}\right)
\end{aligned}
\label{eq:prob}
\end{equation}

If $K \lesssim N $ then the  first two term in the last row of \eqref{eq:prob} cancel to a large extent. If we take the logarithm of \eqref{eq:prob}, we get:
\begin{equation}
\begin{aligned}
\log P(S=n) &\approx n\log{\left(\frac{K-n}{N-n}\right)}+\log{\left(1-\frac{K-n}{N-n}\right)}\\
&\approx n\log(f)+\log(1-f)
\end{aligned}
\label{eq:logprob}
\end{equation}

where $f = (K-n)/(N-n)\simeq K/N$ is approximately the fraction of black balls in the urn. Equation \eqref{eq:logprob} shows that probability of a cascade is exponentially distributed, at least as long as $n \ll N$. Furthermore, as $K$ increases then $f \rightarrow 1$, which implies an uniform distribution.

We propose a different urn model. Let $N_{0}, K_{0}$ and $W$ be the initial number of balls, respectively black and white ones. The procedure of the model is as follows:
\begin{enumerate}
    \item Black balls are drawn with replacement.
    The number of black balls is increased by $W/k$ after each draw. We continue to draw until a white ball is drawn. 
    \item The process repeats with resetting the number of ball to their original values $N = N_{0}, K = K_{0}$.
\end{enumerate}

The rank distribution of the cascade sizes is plotted in Fig. \ref{fig:2}(b) for $k = 1.5$. As we can see, the graph is very well fit by a straight line, indicating that a genuine power law emerges across the entire range of cascade sizes.

At the $i$-th extraction we have the total number of balls $N_{i}$, of black balls $K_{i}$ and white balls $W$. Then:
\begin{equation}
    N_{i} = K_{i} + W
\end{equation}
The urn model is designed such that at each extraction the ratio of $K_{i}$ and $N_{i}$ is approximately given by:
\begin{equation}
    \frac{K_{i}}{N_{i}} = e^{-k/(i+1)}
\end{equation}
Using the previous two equations we obtain that:
\begin{equation}
\begin{aligned}
K_{i} &= \frac{e^{-k/(i+1)}}{1-e^{-k/(i+1)}}W\\
&\simeq \left(\frac{i+1}{k}-1\right)W
\end{aligned}
\end{equation}
The number of black balls is increased after each draw by:
\begin{equation}
K_{i+1}-K_{i} \simeq \frac{W}{k}
\end{equation}
If a constant $C$ number of black balls is added after each draw, then $k = W/C$.

\section{Discussion}

The relevance of this study relies upon two aspects: the increasing importance of understanding statistical laws in complex systems due to the availability of larger datasets \cite{Gerlach2019}, and the presence of power law distributions in various application domains \cite{Reed2001, Pinto2012, Gabaix2016}. The applicability of the models proposed in section 3\ref{subsec:pleq} is restricted to phenomena where time plays a key role in increasing cascade sizes. The more time passes from the initial observation, the greater the chances of seeing a larger cascade (which would place it at the right-end tail of the distribution at the time of observation). These phenomena can appear in a wide variety of disciplinary fields.

The frequency versus the amplitude of earthquakes \cite{Newman2005} and avalanches \cite{Naisbitt2008, Polizzi2021} follows a power law, as well as the distribution of the peak gamma-ray intensity of solar flares \cite{Lu1991}, and the size of moon craters per surface area \cite{Neukum1994}. The largest observed cascade in these examples depends on time, with more extreme examples being discovered the longer the observation record is. 

The social sciences also offer numerous examples of power laws distributions \cite{Turcotte2002} that naturally evolve in time (but whose time dependence has not necessarily been quantified), such as the range of time between two deaths in serial killers' behavioural patterns \cite{SIMKIN2014111}, in narrative structure \cite{Gessey-Jones2020} or in the budget distribution of movies \cite{Bertolotti2022}. In economics, they mainly derive from aggregation and rich-get-richer processes. The existence of power laws in the distribution of wealth and income are known at least from the 19th century \cite{Pareto1896}. Notably, the work of Pareto was the first one to discover the presence of a power law distribution in a social system \cite{Gabaix2016}. Power law distributions also appear in markets. For example, the frequency of firms size, measured by the number of employees, is distributed according to a power law \cite{Axtell2001}. Specifically, the exponent of this power law is approximately $1.06$, which suggests there exists a “Zip’s law” for firms \cite{Gabaix2016}.

Western popular music markets present power law distributions in the lifetime of albums \cite{Schneider2019}. Stock markets exhibited power law distributions since their very beginning \cite{Ribeiro2018}. It has been shown that the number of trades per day, the size of price movements when a large volume of shares are bought or sold, and the number of shares traded per time period are power law distributed with exponents $3$, $0.5$ and $1.5$, respectively \cite{Plerou2005,Bouchaud2009,Kyle2013}. Inter-trading times also exhibit scaling properties consistent with power laws \cite{ivanov2004common}.

Another field in which power law distributions are pervasive is urban studies, where there are many scaling relationships between quantifiable city properties \cite{Wu2019}, such as between the number of gas stations and population levels \cite{Kuhnert2006}, and the distribution of the population of cities in a given geographic area \cite{Gabaix2004}. This latter distribution appears in different geographic areas \cite{Rozenfeld2011} and depends on the granularity of the sampling \cite{Mori2020}. Power law distributions also characterize digital infrastructure. The network analysis of internet topology shows that at the end of the ’90s the degrees of the nodes were already power law distributed \cite{Barabasi1999,Faloutsos1999}. Also, the number of connections to a server in a single day, by a specific subsection of internet users, follows a power law \cite{Adamic2005}.

Furthermore, the academic system generates power laws. As first noted by Price, the number of citations of papers follows a power law distribution \cite{DeSollaPrice1965}. Similar features in academic systems can be observed in recent times, even in specific disciplinary areas. For instance, the cumulative number of citations over time for papers dealing with protein kinases is distributed as a power law. This is a consequence of a phenomenon called the Harlow-Knapp (H-K) effect, which is the propensity of the biomedical and pharmaceutical research communities to concentrate their research on a tiny fraction of the proteome \cite{Grueneberg2008, Fedorov2010}.

In history, the well-known Richardson’s law states that the frequency of wars’ sizes (measured in causalities) are power law distributed. This law is considered one of the few robust statistical regularities in studies of political conflict \cite{Clauset2010}. Nevertheless, some recent studies seem to cast doubt on Richardson’s law, or at least suggest having a greater caution regarding it \cite{Zwetsloot2018}. Another example from history regards the Roman Empire \cite{roman2019growth}, for which the survival time of its emperors is distributed according to a power law \cite{Ramos2021}.

Certain power law relationships do not dependent on the observation period and the proposed models would not apply. For example, in physics, the well-known Stefan–Boltzmann law describes a power law relationship between the total amount of energy radiated per time from a black body due to its temperature \cite{Stefan1879, Boltzmann1884}. Furthermore, in particle physics and astrophysics there are numerous examples of power laws, from the Tully-Fisher relation between a galaxy's luminosity and its rate of rotation \cite{tully1977new}, to the proportionality of the spin and mass of hadrons \cite{chew1962regge} and the density of eigenvalues of the Dirac operator in certain theories \cite{cheng2013scale, del2013large}. Other examples come from biology, where many works describe a scaling relationship between the basal metabolic rate and the body mass of animals \cite{brody1946bioenergetics, benedict1938vital, kleiber1961fire}, such as the well-known Kleiber's law, which indicates that the basal metabolism of mammals increases according to $m^{3/4}$ where $m$ is the overall body mass \cite{Niklas2015}.

Another well known example (with no time dependence) is Zipf’s law, which states that the rank frequency distribution of words in any sufficiently long text is distributed as a power law \cite{Zipf1949}. This law is independent of the language and holds even for artificial languages such as Esperanto \cite{Manaris2006}. While Zipf investigates specifically the distribution of words, a generalization of this law was later developed by Mandelbrot \cite{Mandelbrot1968}.

We end the discussion with a cautionary note on the discovery of power law in data and models. As is well-noted in the literature \cite{Newman2005}, linear relationships in log-log plots in the frequency distributions of observations is insufficient to robustly establish that a power law holds. A more reliable test is to check if the linear relationship holds in log-log plots of rank distributions. As we have shown, the model proposed by Brunk \cite{Brunk2000} does not show power law behaviour. Another example where a power law is claimed to be observed is in the emergence of a scale-free network in systems with bounded rationality \cite{kasthurirathna2015emergence}. As other work has shown \cite{roman2017topology}, the topology is not scale-free but rather of a core-periphery type.
~\\
\section{Conclusion}

In this paper, we have proposed a new mechanism to generate power laws based on Markov models. By analogy with the derivation of the Fokker-Planck equation, we have obtained a master equation for power laws. The result is supported by the fact that the underlying embedded Markov chain has an equilibrium distribution that is a power law and because a simplified version of the equation admits a closed form solution that is a power law.

We proposed three models: a discrete time and discrete space Markov chain with transition probabilities \eqref{eq:pltp}, a continuous time and discrete space Markov chain shown in Fig. \ref{fig:1}(b) and a Markov process (continuum in both time and space), with master equation given by
\eqref{eq:pleq}. A simplified form of \eqref{eq:pleq} is equation \eqref{eq:plsim}, which is valid in the limit of large cascades sizes. Further considerations indicate that an exponent of $k = 2$ gives additional cancellations independent of cascade size. This value of $k$ is consistent with exponents observed in Richardson's law \cite{Zwetsloot2018} and in the net worth of Americans \cite{Newman2005}. Nevertheless, equation \eqref{eq:plsim} has a closed form solution for any exponent value.In addition, the equation provides the time dynamics for the power law relationship along with a natural cut-off size depending on the time horizon.

The stationary solutions for the general equation are given by a first order differential equation, whose solutions have been explored elsewhere \cite{biro2018unidirectional}. However, a stationary solution to the power law equation might not always be well defined, as our time-dependent solution \eqref{eq:csol} for the simplified equation \eqref{eq:plsim} shows (the time derivative is non-zero). The insight we get from the exact time-dependent solution \eqref{eq:fpsol} of the Fokker-Planck equation is that the standard deviation of spatial displacements is proportional to the square root of time, which is characteristic of Brownian motion. Similarly, the time-dependent solution \eqref{eq:csol} gives the largest cascade size we can expect to see is linearly proportional to time.

Based on the insights from the Markov process, we propose a simple urn model that illustrates power law behaviour over the entire range of cascade sizes. The model only considers a constant addition of balls over time and despite the simplicity of the mechanism, a robust power law is observed to emerge. Our results are in contrast to a prior model by Brunk \cite{Brunk2000} which we prove does not show a genuine power law distribution. Finally, we discuss some possible applications of the proposed model to phenomena where power laws are observed and where the duration of the period of observation is important. As far as we know, all the contributions build on existing literature in novel ways.

\begin{acknowledgments}
This work was supported by the Grantham Foundation for the Protection of the Environment. We also thank the World University Network (WUN) for early support of this work in 2014.
\end{acknowledgments}

\bibliography{apssamp}

\end{document}